\documentclass[10pt]{article}
\usepackage[french,english]{babel}
\usepackage{amsmath}
\usepackage{amsfonts}
\usepackage{amssymb}
\usepackage{graphicx}                              
\usepackage{times}
\usepackage{version}
\begin{document}
%
%
October 29th, 2012   \hfill
\vskip 5.8cm
{\baselineskip 12pt
\begin{center}
{\bf UNLOCKING THE STANDARD MODEL}
\end{center}
\begin{center}
{\bf I .\quad  1\ \ GENERATION OF QUARKS . \ SYMMETRIES}
\end{center}
}
\baselineskip 16pt
\vskip .2cm
\centerline{B.~Machet
     \footnote[1]{LPTHE tour 13-14, 4\raise 3pt \hbox{\tiny \`eme} \'etage,
          UPMC Univ Paris 06, BP 126, 4 place Jussieu,
          F-75252 Paris Cedex 05 (France),\\
         Unit\'e Mixte de Recherche UMR 7589 (CNRS / UPMC Univ Paris 06)}
    \footnote[2]{machet@lpthe.jussieu.fr}
     }
\vskip 1cm

{\bf Abstract:} A very specific two-Higgs-doublet extension of the
Glashow-Salam-Weinberg  model for one generation of quarks is advocated
for, in which the two
doublets are parity transformed of each other and both isomorphic to the Higgs
doublet of the Standard Model.  The chiral group $U(2)_L \times U(2)_R$
gets broken down to $U(1) \times U(1)_{em}$. In there, the first diagonal
$U(1)$ is directly connected to parity through the $U(1)_L \times U(1)_R$
algebra.  Both chiral and weak symmetry breaking
can be accounted for, together with their relevant degrees of freedom.
The two Higgs doublets are demonstrated to be in one-to-one correspondence
with bilinear quark operators.

\smallskip



\section{Introduction}
\label{section:intro}

It is well known that the genuine Glashow-Salam-Weinberg (GSW) model
\cite{GSW} cannot account for both  chiral and weak symmetry breaking.
For composite Higgses, the failure of the most simple scheme of
dynamical symmetry breaking
 bred ``technicolor'' models \cite{Susskind} \cite{FahriJackiw} \cite{technicolor},
in which at least one higher scale and extra
heavy fermions are needed. They unfortunately face themselves many problems.
Guided by a long quest \cite{Machet} for all complex doublets isomorphic to the Higgs
doublet of the Standard Model, I propose here a very simple and natural
two-Higgs-doublet extension in which, in particular, parity is restored to
the primary role that it is expected to play and where the
issues evoked above are solved.
Last, an isomorphism is
demonstrated which connects the two proposed Higgs doublets and bilinear
quark-antiquark operators.
This works therefore achieves a synthesis of two-Higgs-doublet models
and dynamical symmetry breaking.
The arguments that I present are essentially based on elementary
considerations concerning symmetries.
For the sake of simplicity, only the case of one generation of quarks
is dealt with here.

\section{The Glashow-Salam-Weinberg model and its single Higgs doublet}
\label{section:GSW}

A $SU(2)_L$ transformation we shall write
\begin{equation}
{\cal U}_L = e^{-i\alpha_i T^i_L},\quad i=1,2,3.
\label{eq:groupL}
\end{equation}
Eq.~(\ref{eq:groupL}) shows that we adopt here the convention of physicists
and consider the hermitian $T^i$'s as the generators of $SU(2)_L$.
So defined, they satisfy the commutation relations
\begin{equation}
[T^i_L,T^j_L]= i\,\epsilon_{ijk}T^k_L,
\label{eq:commut}
\end{equation}
and they write
\begin{equation}
\vec T_L = \frac12 \vec \tau,
\label{eq:gen1}
\end{equation}
where the $\vec\tau$ are the Pauli matrices
\begin{equation}
\tau^1=\left(\begin{array}{rr}0 & 1\cr 1 & 0\end{array}\right),\quad
\tau^2=\left(\begin{array}{rr}0 & -i\cr i & 0\end{array}\right),\quad
\tau^3=\left(\begin{array}{rr}1 & 0\cr 0 & -1\end{array}\right).
\label{eq:Pauli}
\end{equation}
The $T^i_L$'s are accordingly
\begin{equation}
T^3_L=\frac12 \left(\begin{array}{rr} 1 & 0 \cr 0 & -1 \end{array}\right),\quad
T^+_L = T^1_L+iT^2_L= 
\left(\begin{array}{cc} 0 & 1 \cr 0 & 0
\end{array}\right),\quad
T^-_L = T^1_L-iT^2_L= 
 \left(\begin{array}{cc} 0 & 0 \cr 1 & 0 \end{array}\right).
\label{eq:gen2}
\end{equation}
With respect to $SU(2)_L$, left-handed flavor
 fermions are cast into doublets
$\left(\begin{array}{c}u_L\cr d_L\end{array}\right)$,
while right-handed fermions are singlets. These doublets belong
to the fundamental representation of $SU(2)_L$ and the generators $T^i_L$ act on
them accordingly by
\begin{equation}
T^i_L.\left(\begin{array}{c} u_L\cr d_L\end{array}\right) = 
T^i_L\left(\begin{array}{c} u_L\cr d_L\end{array}\right).
\end{equation}
The Higgs doublet $H$ of the GSW model is a complex doublet
\begin{equation}
H = \frac{1}{\sqrt{2}}\left(\begin{array}{c}  \chi^1 +i \chi^2 \cr \chi^0 - ik^3
                       \end{array}\right),\quad 
<\chi^0>=v \Leftrightarrow 
<H> = \frac{1}{\sqrt{2}}\left(\begin{array}{c} 0 \cr v\end{array}\right),
\label{eq:doublet}
\end{equation}     
also in the fundamental representation of $SU(2)_L$. It is built with four
 real scalar fields
$\chi_0=\frac{v}{\sqrt{2}}+\chi, \chi^1,\chi^2,k^3\equiv -i\chi^3$.
We have set $\chi^3=ik^3$ in (\ref{eq:doublet}) to emphasize that it is complex.
The action of $T^i_L$ on $H$ writes
\begin{equation}
T^i_L\,.\,H = T^i_L H.
\label{eq:action}
\end{equation}

$H$ is used to give a mass to the three gauge bosons and also, through
Yukawa couplings, to the $d$-type quarks. To give a mass to $u$-type quarks,
a second (non-independent) complex doublet $\tilde H$
\begin{equation}
\tilde H = i\tau^2 H^\ast 
= \frac{1}{\sqrt{2}}\left(\begin{array}{c}  \chi^0 + ik^3 \cr
-(\chi^1-i\chi^2)
                       \end{array}\right),\ 
<\tilde H> = \frac{1}{\sqrt{2}}\left(\begin{array}{c} v \cr
0\end{array}\right),
\label{eq:doublet2}
\end{equation}     
is used which is also in the fundamental representation of $SU(2)_L$. It 
has  the same law of transformation (\ref{eq:action})
\footnote{
If one considers $\tilde H = \beta_j T^j H^*$ and request that, by a
transformation $e^{-i\alpha_i T^i_L}$, $\delta \tilde H = -i\alpha_iT^i_L.\tilde
H = -i\alpha_iT^i_L \tilde H$, one gets the condition $\beta_1=0=\beta_3$.
$\beta_2$ is undetermined and can be taken to be $\beta_2=2i$.}.

We define the transformed $T^i_L.\chi^\alpha, i=1,2,3,\alpha=0,1,2,3$ of
the components $\chi^\alpha$ by
\begin{equation}
T^i_L.H = \frac{1}{\sqrt{2}}\left(\begin{array}{c} 
T^i_L.\chi^1 + i T^i_L.\chi^2\cr
T^i_L.\chi^0 - T^i_L.\chi^3\end{array}\right)
\label{eq:transcomp1}
\end{equation}
and the same for $\tilde H$.
The law of transformation (\ref{eq:action}), when applied to both $H$ and $\tilde
H$, is equivalent to
\medskip

\vbox{
\begin{equation}
\begin{array}{lll}
 T^1_L\,.\,\chi^0=+\frac{i}{2}\,\chi^2,&
T^2_L\,.\,\chi^0=+\frac{i}{2}\,\chi^1,&
T^3_L\,.\,\chi^0=+\frac12\, \chi^3,\cr
 T^1_L\,.\,\chi^1=-\frac{1}{2}\,\chi^3,&
T^2_L\,.\,\chi^1=-\frac{i}{2}\,\chi^0,&
T^3_L\,.\,\chi^1=+\frac{i}{2}\, h^2,\cr
 T^1_L\,.\,\chi^2=-\frac{i}{2}\,\chi^0,&
T^2_L\,.\,\chi^2=+\frac{1}{2}\,\chi^3,&
T^3_L\,.\,\chi^2=-\frac{i}{2}\, \chi^1,\cr
 T^1_L\,.\,\chi^3=-\frac{1}{2}\,\chi^1,&
T^2_L\,.\,\chi^3=+\frac{1}{2}\,\chi^2,&
T^3_L\,.\,\chi^3=+\frac{1}{2}\, \chi^0.
\end{array}
\label{eq:rule}
\end{equation}
}

Let us make the following substitutions
\begin{equation}
\chi^0 \to -h^3,\quad \chi^1\to h^1, \quad \chi^2\to -h^2,\quad
\chi^3\to h^0
\label{eq:subst}
\end{equation}
such that, in terms of the $h^i$'s, $H$ and $\tilde H$ write
\begin{equation}
H = \left(\begin{array}{c}h^1-ih^2 \cr -(h^0+h^3) 
\end{array}\right), \quad
\tilde H = \left(\begin{array}{c}h^0-h^3 \cr -(h^1 + ih^2) 
\end{array}\right).
\label{eq:H1d}
\end{equation}
Then, the laws of transformation (\ref{eq:rule}) rewrite simply
\begin{equation}
\boxed{
\begin{array}{rcl}
T^i_L\,.\,h^j&=&-\frac{1}{2}\left(i\,\epsilon_{ijk}h^k +
\delta_{ij}\,h^0\right)\cr
T^i_L\,.\,h^0 &=& -\frac{1}{2}\, h^i
\end{array}}
\label{eq:ruleL}
\end{equation}
which is our main formula.

Acting a second time on the $\chi^\alpha$ according to the rules
(\ref{eq:rule}) yields
\begin{equation}
T^i_L.(T^j_L\,.\,\chi^\alpha) - T^j_L.(T^i_L\,.\,\chi^\alpha)
 = -i\,\epsilon_{ijk} T^k_L\,.\,\chi^\alpha = -[T^i_L,T^j_L].\chi^\alpha,
\quad \alpha =0,1,2,3.
\label{eq:lie}
\end{equation}

After (\ref{eq:transcomp1}), it is natural to define $(T^i_L
T^j_L).\chi^\alpha$ by
\begin{equation}
(T^i_LT^j_L).H = \frac{1}{\sqrt{2}}\left(\begin{array}{c} 
(T^i_LT^j_L).\chi^1 + i (T^i_LT^j_L).\chi^2\cr
(T^i_LT^j_L).\chi^0 - (T^i_LT^j_L).\chi^3\end{array}\right),
\label{transcomp2}
\end{equation}
in which $T^i_LT^j_L$ stands for the product of the two corresponding
matrices. Since they satisfy (\ref{eq:commut}) and by using (\ref{eq:rule}), one gets
straightforwardly
\footnote{Eqs.~(\ref{eq:lie}) and (\ref{eq:liebis}) require
\begin{equation}
(T^i_L T^j_L).\chi^\alpha = -T^i_L.(T^j_L.\chi^\alpha).
\label{eq:sign}
\end{equation}
It is easy to trace the origin of the ``$-$'' sign in (\ref{eq:sign}) back to
the definition (\ref{eq:groupL}) and (\ref{eq:gen1})
 of the generators of $SU(2)$ as
hermitian matrices. If one instead defines a group transformation without the
``$i$'' in the exponential, which yields anti-hermitian generators, one gets
identity of $(T^i_L T^j_L).\chi^\alpha$ and $T^i_L.(T^j_L.\chi^\alpha)$.
The ``$i$'' in the r.h.s. of the commutation relation (\ref{eq:commut}) is
 then also canceled.}
\begin{equation}
[T^i_L,T^j_L].\chi^\alpha = +i\,\epsilon_{ijk}T^k_L.\chi^\alpha, \quad
\alpha=0,1,2,3.
\label{eq:liebis}
\end{equation}

\section{Its two-Higgs-doublet avatar}
\label{section:2hdm}

In analogy with (\ref{eq:H1d}), we shall introduce two complex $SU(2)_L$
 Higgs doublets $H$ and $K$ 
\begin{equation}
K = \left(\begin{array}{c}{\mathfrak p}^1-i{\mathfrak p}^2 \cr -({\mathfrak
s}^0+{\mathfrak p}^3) 
\end{array}\right)
\label{eq:Kd}
\end{equation}
and
\begin{equation}
H = \left(\begin{array}{c}{\mathfrak s}^1-i{\mathfrak s}^2 \cr -({\mathfrak
p}^0+{\mathfrak s}^3) 
\end{array}\right)
\label{eq:Hd}
\end{equation}
in which  we defined
${\mathfrak p}^\pm = {\mathfrak p}^1 \pm i{\mathfrak p}^2$
and
${\mathfrak s}^\pm = {\mathfrak s}^1 \pm i{\mathfrak s}^2$.
We shall also indifferently speak of $K$ as  of the quadruplet 
\begin{equation}
K=({\mathfrak s}^0, {\mathfrak p}^3, {\mathfrak p}^+, {\mathfrak p}^-),
\label{eq:Kq}
\end{equation}
and of $H$  as of the quadruplet
\begin{equation}
H=({\mathfrak p}^0, {\mathfrak s}^3, {\mathfrak s}^+, {\mathfrak s}^-).
\label{eq:Hq}
\end{equation}
$K$ and $H$ have the same laws of transformations (\ref{eq:ruleL}) by
$SU(2)_L$, in which, now $h$ stands generically for $\mathfrak s$ or $\mathfrak p$,
Each component of the type $\mathfrak s$ is hereafter considered as a scalar
and each $\mathfrak p$ as a pseudoscalar, such that $H$ and $K$ are parity
transformed of each other.

$K$ and $H$ are also stable by $SU(2)_R$, the generators $T^i_R$ of which
being defined like in (\ref{eq:gen2}), and transform according to
\begin{equation}
\boxed{\begin{array}{rcl}
T^i_R\,.\,h^j&=&-\frac{1}{2}\left(i\,\epsilon_{ijk}h^k -
\delta_{ij}\,h^0\right)\cr
T^i_R\,.\,h^0 &=& +\frac{1}{2}\, h^i\end{array}}
\label{eq:ruleR}
\end{equation}
Since $h^0$ and $\vec h$ have opposite parity,
(\ref{eq:ruleL}) and (\ref{eq:ruleR}) mix scalars and pseudoscalars.

Last, let us define the action of the generators $I_L$ and $I_R$ of
$U(1)_L$ and $U(1)_R$ on $K$ and $H$ by (this is true component by
component)
\begin{equation}
\boxed{
I_L.K = -H, \quad I_L.H=-K,\quad I_R.K=H,\quad I_R.H=K
}
\label{eq:par}
\end{equation}
they simply swap parity with the appropriate signs.

From eqs.~(\ref{eq:ruleL}), (\ref{eq:ruleR}) and (\ref{eq:par}), one deduces the
laws of transformations of $K$ and $H$ by the diagonal $U(2)$
\vbox{
\begin{eqnarray}
T^i\,.\,h^j&=&-i\,\epsilon_{ijk}h^k,\cr
T^i\,.\,h^0 &=& 0,\cr
I\,.\,h^{0,i} &=& 0.
\label{eq:ruleD}
\end{eqnarray}
}

$K$ and $H$ thus
decompose into a singlet $h^0$ + a triplet $\vec h$ of the diagonal $SU(2)$.
They are stable by $SU(2)_L$ and $SU(2)_R$ but not  by $U(1)_L$ nor by
$U(1)_R$.  They stay unchanged when acted upon by the diagonal $U(1)$.
The union $K \cup H$ is stable by the whole chiral group $U(2)_L \times
U(2)_R$.

It is instructive to represent the $\vec T_L$, $\vec T_R$ generators, and
also the $\vec T$ of the diagonal $SU(2)_V$ symmetry in the basis
the elements of which are the four entries $(h^0, h^3, h^+,
h^-)$  of $K$ and $H$ (see appendix \ref{section:4basis} for explicit
formul{\ae}).
The $SU(2)_{L,R}$ generators  satisfy the commutation relations
\begin{equation}
[T^+_{L,R},T^-_{L,R}]=-2T^3_{L,R},\quad [T^3_{L,R}, T^+_{L,R}]=-T^+_{L,R},
\quad [T^3_{L,R},T^-_{L,R}]=T^-_{L,R},
\label{eq:comLR}
\end{equation}
which become the customary $SU(2)$ commutation relations when the roles of $T^+$ and
$T^-$, or of $h^+$ and $h^-$, are swapped. Left and right generators of
course commute. They also  satisfy the anticommutation relations
\begin{equation}
 \{T^i_{L,R}, T^j_{L,R}\} = \frac12\,\delta_{ij} { I}_{L,R},
\end{equation}
which makes them like $4\times 4$ Pauli matrices.
The three generators of the diagonal $SU(2)$ have the same commutation
relation but no peculiar anticommutation relation.

$K$ and $H$ are not eigenvectors of $\vec T_L$, $\vec T_R$ or $\vec T$.
Calculating the eigenvectors of the $SU(2)_L$ generators
 $\vec T_L$ and $\vec T_R$ given by (\ref{eq:t3l}) and (\ref{eq:t3r}) leads 
 to 4-vectors gathering the two doublets $\left(\begin{array}{c}h^-\cr
-(h^0+h^3)\end{array}\right)$ and $\left(\begin{array}{c}
h^0-h^3\cr -h^+\end{array}\right)$ isomorphic respectively to $H$
and $\tilde H$ (\ref{eq:H1d}) of the GSW model.

In this 4-basis, the third generator $T^3=T^3_L + T^3_R$ of the diagonal
$SU(2)$ symmetry coincides with the electric charge
\footnote{That the electric charge is the 3rd generator of an $SU(2)$ group
is a prerequisite for charge quantization.}
$Q$,
with eigenvalues $0$ (twice), $+1$ and $-1$. Then, the Gell-Mann-Nishijima
relation $Y =  Q - T^3_L$  identifies the weak hypercharge $ Y$ with $T^3_R$.

\section{Symmetries and their breaking}
\label{section:ssb}

In association with a suitable potential, the gauge symmetry gets
spontaneously broken by the presence of non-vanishing vacuum expectation
value(s) (VEV(s)). These VEV's should be electrically neutral. Secondly,
even if, since we deal with a parity violating theory, pseudoscalar fields
can  also be expected to ``condense'' in the vacuum,
such VEV's can reasonably be expected only at higher order such
that, classically, it is legitimate to restrict to scalar VEV's. This leaves
the two ${\mathfrak s}^0$ in $K$ and ${\mathfrak s}^3$ in $H$. We shall write
accordingly
\begin{equation}
{\mathfrak s}^0 = \frac{v}{\sqrt{2}} + s, \quad
{\mathfrak s}^3 = \frac{\sigma}{\sqrt{2}} + \xi,
\label{eq:vev1}
\end{equation}
such that
\footnote{The sign of the VEV's is not relevant, neither for the masses
of the gauge bosons, which depend on their squares, nor for the fermions
since, for
example, the sign of the Yukawa couplings can always be adapted.}
\begin{equation}
<K> = -\frac{1}{\sqrt{2}}\left(\begin{array}{c} 0 \cr v \end{array}\right), 
\quad
<H> = -\frac{1}{\sqrt{2}}\left(\begin{array}{c} 0 \cr \sigma
\end{array}\right).
\label{eq:vev2} 
\end{equation}

Let us first investigate the breaking of the chiral group $U(2)_L \times U(2)_R$.
Two $U(1)$ groups are left unbroken by the vacuum. The first is the
diagonal $U(1)$, which is the multiplication of fermions by a phase. We have seen that
its generator, the unit matrix, is the sum of $I_L$ and $I_R$
which swap parities of the scalar fields. The second is
the electromagnetic $U(1)_{em}$.
Indeed, only $T^3$, that is, the electric charge $Q$, gives $0$
when acting on ${\mathfrak s}^0$ and ${\mathfrak s}^3$. 
$U(2)_L \times U(2)_R$ gets accordingly spontaneously broken down to $U(1)
\times U(1)_{em}$, such that  six Goldstone bosons are expected.
Three degrees of freedom should become the three longitudinal
$\vec W_\parallel$.  The three
others are expected to be physical particles and to acquire mass by a
``soft'' breaking of $U(2)_L \times U(2)_R$. This role is held by the
$SU(2)_L$ invariant Yukawa couplings (we shall study them more at length in
a subsequent work \cite{Machet2}
 and only give a few remarks at the end of this section).

The weak group $SU(2)_L$ has been considered {\em de facto} 
as a subgroup of the chiral group $U(2)_L \times U(2)_R$.
This can always be done
\footnote{and it can even always be done when going to $N$
generations and to the chiral group $U(2N)_L\times U(2N)_R$.
The embedding only needs to be suitably chosen so as to match the weak
Lagrangian and dictates the way chiral and weak symmetries get entangled
\cite{Machet}.} and creates connections among Goldstone
bosons. One among the four Goldstones (two charged and two neutral)
  which arise from
the  breaking of $U(2)_L \times U(2)_R$ down to its diagonal $U(2)$
subgroup is identical to the neutral  Goldstone
 of the breaking of $SU(2)_L$ (which generates three Goldstones) and
gets accordingly eaten by the  massive $W^3$.
The spectrum that results is therefore composed of the three
massive $\vec W$'s, of the three remaining physical
(pseudo)Goldstone bosons of the breaking $U(2)_L \times U(2)_R \to U(2)$
(which are also those of the breaking of $SU(2)_L \times SU(2)_R$ down to
$SU(2)$), and of two Higgs bosons.

Since, for one generation of 
fermions, there is no distinction between the diagonal $SU(2)$ and the
$SU(2)$ of flavor, the pseudoscalar triplet in $K$ cannot but be  ``pion-like''
and the pseudoscalar singlet in $H$ be  ``$\eta$-like''  (it is similar to
$\eta$ for one generation but can be the pseudoscalar singlet or another
combination for more generations). This degree of freedom,
which disappears to the benefit of the  massive  neutral $W^3$,
is presumably the one that plays a dual role with respect to the chiral
and weak symmetry breakings.

The goal being to build a spontaneously broken $SU(2)_L$
 theory of weak interactions, let us make a few more remarks concerning the
breaking of $SU(2)_L$.
All generators $T^i_L$ acting non-trivially on ${\mathfrak s}^0$ (see for
example appendix \ref{section:4basis}),
$SU(2)_L$ gets fully broken by $<K>\not=0$, which yields
three Goldstone bosons inside $K$.  In the genuine GSW model, they  would become the
three longitudinal $\vec W_\parallel$'s. The situation is now changed
as $K$ gets instead connected to chiral breaking. This is where
$H$ enters the game since $SU(2)_L$ gets also fully broken by
 $<{\mathfrak s}^3>\not=0$, which  generates the largest part of the $\vec W$ mass.
Note that the  same argumentation can be applied to $SU(2)_R$ since $K$ and $H$ are
also stable multiplets of $SU(2)_R$.

These considerations fix the three
Goldstones $({\mathfrak p}, {\mathfrak s}^+, {\mathfrak s}^-)$ in $H$, two
charged scalars and one neutral pseudoscalar,
as the ones doomed to become the three longitudinal $W_\parallel$'s.
This  establishes $H$ as the Higgs multiplet the closest to that of the
GSW model and $K$ as the additional ``chiral'' multiplet.
The three $\vec{\mathfrak p}$
in there are  pion-like, and $s$ is a second Higgs boson.

Would parity be unbroken, the two Higgs multiplets $H$ and $K$ would be
equivalent. This would entail in particular that ${\mathfrak s}^0$
and ${\mathfrak p}^0$ have identical VEV's. This is not the case.
The two VEV's $v$ and $\sigma$ do not belong
to fields that are parity transformed of each other but to the 
neutral scalars ${\mathfrak s}^0$ and ${\mathfrak s}^3$. They
are independent parameters, controlling respectively  the
chiral and weak symmetry breaking.
As far as  the diagonal $SU(2)$ is concerned, it gets broken down to its
electromagnetic $U(1)_{em}$ subgroup.

Soft chiral breaking is expected to provide  low masses for the three
$\vec{\mathfrak p}$. This process is achieved through $SU(2)_L$ invariant Yukawa
couplings of fermions to both Higgs multiplets $K$ and $H$.
At high energy $(m_W)$ these are standard renormalizable couplings between scalar
fields and two fermions.  At low energy, they can be rewritten (bosonised)  by using
 the Partially Conserved Axial Current (PCAC) hypothesis
\cite{Dashen}\cite{Lee}\cite{dAFFR}. So doing, they yield \cite{Machet2}
in particular terms which are quadratic in the three $\vec
{\mathfrak p}$ Goldstones and  which match pion-like mass terms
in agreement with the Gell-Mann-Oakes-Renner relation
\cite{GMOR}\cite{dAFFR}.

The situation is therefore different from when one sticks to a unique
Higgs doublet in that we have enough degrees of freedom
to accommodate for both chiral and weak physics of one generation
of quarks. 

This pattern of symmetry breaking  also constrains the quartic
scalar potential $V(H,K)$ that one introduces as a
spontaneous-symmetry-breaking tool. Since the three
$\vec{\mathfrak p}$ in $K$ should match the three Goldstones of the
chiral breaking $SU(2)_L \times SU(2)_R \to SU(2)$,
no mass difference between the neutral and charged components should
be generated in $V$. Other constraints arise from  the requirement that
scalar-pseudoscalar transitions should also be avoided at this
(classical) level. Together, they largely simplify the expression of
$V(H,K)$, which also receives, at low energy, additional contributions from
the bosonised Yukawa couplings.

\section{The isomorphism between the two Higgs doublets
and bilinear quark operators}
\label{section:analogy}

Like for $SU(2)_L$, we define a $SU(2)_R$ transformation by
\begin{equation}
{\cal U}_R = e^{-i\beta_j T^j_L},\quad j=1,2,3,
\label{eq:groupR}
\end{equation}
in which the generators $T^j_R$ are given by the same hermitian matrices as in
(\ref{eq:gen2}).
The $U(2)_L \times U(2)_R$ algebra is completed by the two generators $I_L$
and $I_R$, each one being represented by the unit $2\times 2$ matrix.

Any quark bilinear car be represented as $\bar \psi
{\mathbb M}\psi$ or $\bar\psi{\mathbb M}\gamma_5\psi$, where $\mathbb M$ is
also a $2\times 2$ matrix, or, equivalently, as ``even'' and ``odd''
composite operators $\bar\psi \frac{1+\gamma_5}{2}{\mathbb M}\psi$ and
$\bar\psi\frac{1-\gamma_5}{2}{\mathbb M}\psi$.

The laws of transformations of (even and odd) fermion bilinears
by an element ${\cal U}={\cal U}_L \times {\cal U}_R$ of
the chiral group are defined as follows: 

\vbox{
\begin{eqnarray}
({\cal U}_L \times {\cal U}_R)\,.\,\bar\psi\frac{1+\gamma_5}{2}{\mathbb M}\psi
&=& \bar \psi\; {\cal U}_L^{-1}\,{\mathbb M\;{\cal U}_R\;\frac{1+\gamma_5}{2}}\psi,\cr
&& \cr
({\cal U}_L \times {\cal U}_R)\,.\,\bar\psi\frac{1-\gamma_5}{2}{\mathbb M}\psi
&=& \bar \psi\; {\cal U}_R^{-1}\,{\mathbb M\;{\cal U}_L\;\frac{1-\gamma_5}{2}}\psi,
\label{eq:group}
\end{eqnarray}
}

which gives, by expanding ${\cal U}_L = 1-i\beta_j{T}^j_L + \ldots$ and
${\cal U}_R = 1-i\kappa_j{T}^j_R + \ldots$

\vbox{
\begin{eqnarray}
{T}^j_L\,.\,\bar\psi\frac{1+\gamma_5}{2}{\mathbb M}\psi &=&
 -\,\bar\psi\, {T}^j{\mathbb M}\, \frac{1+\gamma_5}{2}\psi,\cr
&& \cr
{T}^j_L\,.\,\bar\psi\frac{1-\gamma_5}{2}{\mathbb M}\psi &=&
+\,\bar\psi\, {\mathbb M}{T}^j\, \frac{1-\gamma_5}{2}\psi,\cr
&& \cr
{T}^j_R\,.\,\bar\psi\frac{1+\gamma_5}{2}{\mathbb M}\psi &=&
+\,\bar\psi\, {\mathbb M}{T}^j\, \frac{1+\gamma_5}{2}\psi,\cr
&& \cr
{T}^j_R\,.\,\bar\psi\frac{1-\gamma_5}{2}{\mathbb M}\psi &=&
 -\,\bar\psi\, {T}^j{\mathbb M}\, \frac{1-\gamma_5}{2}\psi.
\label{eq:trans}
\end{eqnarray}
}

The ${T}^j$'s are seen to simply act by left- or
right-multiplication on the matrix $\mathbb M$.
Eqs.~(\ref{eq:trans}) give, for scalar and pseudoscalar bilinears

\vbox{
\begin{eqnarray}
{ T}^j_L\,.\,\bar\psi{\mathbb M} \psi &=&
-\frac12\,\left(\bar\psi\,[{ T}^j,{\mathbb M}]\,\psi
                 +\bar\psi\, \{{ T}^j,{\mathbb M}\}\,\gamma_5\psi\right),\cr
&& \cr
{ T}^j_L\,.\,\bar\psi{\mathbb M}\gamma_5 \psi &=&
-\frac12\,\left(\bar\psi\,[{ T}^j,{\mathbb
M}]\,\gamma_5\psi
                 +\bar\psi\, \{{ T}^j,{\mathbb M}\}\,\psi\right),\cr
&& \cr
{ T}^j_R\,.\,\bar\psi{\mathbb M} \psi &=&
-\frac12\,\left(\bar\psi\,[{ T}^j,{\mathbb M}]\,\psi
                 -\bar\psi\, \{{ T}^j,{\mathbb M}\}\,\gamma_5\psi\right),\cr
&& \cr
{ T}^j_R\,.\,\bar\psi{\mathbb M}\gamma_5 \psi &=&
-\frac12\,\left(\bar\psi\,[{ T}^j,{\mathbb
M}]\,\gamma_5\psi
                 -\bar\psi\, \{{ T}^j,{\mathbb M}\}\,\psi\right),
\label{eq:trans2}
\end{eqnarray}
}

in which $\{,\}$ stands for the anticommutator of two matrices.
Then, by using commutation and anticommutation relations of the Pauli
matrices, one notices that the set of $2\times 2$ matrices $(I, \vec T)$ is
stable by these two operations. One then gets from (\ref{eq:trans2})

\vbox{
\begin{eqnarray}
T^i_L\,.\, \bar\psi { I} \psi &=& -\frac12\, \bar \psi \gamma_5 2T^i \psi,\cr
&&\cr
T^i_L\,.\,\bar \psi\gamma_5 2T^j\psi &=& -\frac12\, \left( i\,\epsilon_{ijk}\; \bar\psi
\gamma_5 2T^k\psi + \delta_{ij}\; \bar\psi { I} \psi\right),
\label{eq:law1}
\end{eqnarray}
}

and

\vbox{
\begin{eqnarray}
T^i_L\,.\, \bar\psi \gamma_5{ I} \psi &=& -\frac12\, \bar \psi  2T^i \psi,\cr
&&\cr
T^i_L\,.\,\bar \psi 2T^j\psi &=& -\frac12\, \left( i\,\epsilon_{ijk}\; \bar\psi
2T^k\psi + \delta_{ij}\; \bar\psi \gamma_5{ I} \psi\right),
\label{eq:law2}
\end{eqnarray}
}

which show that the two quadruplets, which are parity-transformed of each
other
\begin{equation}
\Phi=(\phi^0,\vec\phi)=\bar\psi\left(
{ I}, 2\gamma_5 \vec T \right)\psi
\label{eq:phi}
\end{equation}
and
\begin{equation}
\Xi=(\xi^0,\vec\xi)=\bar\psi\left(
\gamma_5{ I}, 2\vec T\right)\psi
\label{eq:xi}
\end{equation}
are stable by $SU(2)_L$ and have the following laws of transformation
(we write them for $\Phi$, the ones for $\Xi$ are identical)

\vbox{
\begin{equation}
\boxed{
\begin{array}{rcl}
T^i_L\,.\,\phi^j&=&-\frac{1}{2}\,\left(i\,\epsilon_{ijk}\phi^k +
\delta_{ij}\,\phi^0\right)\cr
T^i_L\,.\,\phi^0 &=& -\frac{1}{2}\, \phi^i\end{array}}
\label{eq:rule2}
\end{equation}
}

The same considerations can be applied to $SU(2)_R$, which leads to
\medskip

\vbox{
\begin{equation}
\boxed{
\begin{array}{rcl}
T^i_R\,.\,\phi^j&=&-\frac{1}{2}\,\left(i\,\epsilon_{ijk}\phi^k -
\delta_{ij}\,\phi^0\right)\cr
T^i_R\,.\,\phi^0 &=& +\frac{1}{2}\, \phi^i\end{array}}
\label{eq:rule3}
\end{equation}
}

Acting with the generators $I_L$ and $I_R$ of $U(1)_L$ and $U(1)_R$
on any scalar $S=\bar\psi{\mathbb M}\psi$ of pseudoscalar $P=\bar\psi{\mathbb
M}\gamma_5\psi$ according to (\ref{eq:trans2}) yields
\footnote{At the group level, $S$ and $P$ transform according to
\begin{equation}
\begin{array}{c}
e^{-i\alpha I_L}\,.\,(S,P)= \cos\alpha (S,P) + i\sin\alpha(P,S),\cr
e^{-i\alpha I_R}\,.\,(S,P)= \cos\alpha (S,P) - i\sin\alpha(P,S).
\end{array}
\end{equation}
As expected since the phases of $\psi$ and $\bar\psi$ cancel,
by a transformation $e^{-i\alpha I}$ of
the diagonal $U(1)$, any $S$ and $P$ is left invariant.}
\begin{equation}
\boxed{
I_L\,.\,S=-P,\quad I_L\,.\,P=-S,\quad I_R\,.\,S=P,\quad I_R\,.\,P=S
}
\label{eq:parity}
\end{equation}
Eqs.~(\ref{eq:rule2}), (\ref{eq:rule3}) and (\ref{eq:parity},
 being identical to (\ref{eq:ruleL}), (\ref{eq:ruleR}) and (\ref{eq:par}),
 establish the isomorphism between $K$ and $H$
and the composite multiplets $\Xi$ and $\Phi$.
It is completed by going from $\Phi$ and $\Xi$ to multiplets $\mathfrak K$
and $\mathfrak H$ which have dimension $[mass]$ like $K$ and $H$:

\vbox{
\begin{eqnarray}
{\mathfrak K}
=\frac{1}{\sqrt{2}}\frac{v}{\mu^3}\left(\begin{array}{c}
\phi^1-i\phi^2 \cr
-(\phi^0+\phi^3)\end{array}\right)
&=&\frac{v\sqrt{2}}{\mu^3}\left(\begin{array}{c}
\bar d \gamma_5 u \cr
-\frac12(\bar u u + \bar d d) -\frac12(\bar u\gamma_5 u - \bar d \gamma_5 d)
\end{array}\right)
\equiv\left(\begin{array}{c} {\mathfrak k}^1-i{\mathfrak
k}^2\cr-({\mathfrak k}^0+{\mathfrak k}^3)\end{array}\right),\cr
%
&& \cr
<\bar u u + \bar d d> &=& \mu^3,\cr
&& \cr
{\mathfrak H}
=\frac{1}{\sqrt{2}}\frac{\sigma}{\nu^3}\left(\begin{array}{c}
\xi^1-i\xi^2\cr
-(\xi^0+\xi^3)\end{array}\right)
&=&\frac{\sigma\sqrt{2}}{\nu^3}\left(\begin{array}{c}
\bar d  u \cr
-\frac12(\bar u\gamma_5 u + \bar d\gamma_5 d) -\frac12(\bar u u - \bar d  d)
\end{array}\right)
\equiv\left(\begin{array}{c} {\mathfrak h}^1-i{\mathfrak
h}^2\cr-({\mathfrak h}^0+{\mathfrak h}^3)\end{array}\right),\cr
%
&& \cr
<\bar u u - \bar d d> &=& \nu^3.\cr
&&
\label{eq:compdoub}
\end{eqnarray}
}

In particular, as far as the Higgs bosons are concerned, the isomorphism
writes
\begin{equation}
{\mathfrak s}^0 \leftrightarrow \frac{v}{\sqrt{2}\mu^3}\,(\bar u u + \bar
d d),\quad
{\mathfrak s}^3 \leftrightarrow \frac{\sigma}{\sqrt{2}\nu^3}\,(\bar u u - \bar
d d).
\end{equation}
The components of $\mathfrak K$ and $\mathfrak H$ have of course the same
laws of transformations (\ref{eq:rule2}) and (\ref{eq:rule3}) as the ones of $\Phi$
and $\Xi$.
In this framework, the VEV's of the scalar bilinear fermion operators
$\bar u u$ and $\bar d d$ act
as catalysts for both chiral and weak symmetry breaking.

\section{Conclusion and prospects}
\label{section:conclusion}

We advocate for a minimal extension of the Glashow-Salam-Weinberg model for one
generation of fermions which simply endows it with two Higgs multiplets instead
of one, the two of them being  parity transformed of each other
\footnote{This last property distinguishes it 
 from all two-Higgs-doublet extensions that we are aware of \cite{BFLRSS}\cite{HHG}
\cite{DiazSanchez}.}.
This procedure unlocks the Standard Model in the sense that
both chiral and weak symmetry breaking can now be accounted for and that enough
degrees of freedom become available to describe the two corresponding
 scales of physics: a first Higgs multiplet carries the three would-be longitudinal
$\vec W_\parallel$ + a scalar Higgs boson which is very much
``standard-like'', and a second  multiplet carries the three (pseudo)Goldstones
of the broken $SU(2)_L\times SU(2)_R$ chiral symmetry into the diagonal
$SU(2)$ + an additional Higgs boson $s$. The six
Goldstones  can be traced back to the breaking of $U(2)_L \times U(2)_R$ down to
$U(1) \times U(1)_{em}$. The first $U(1)$, which is the diagonal subgroup of
$U(1)_L \times U(1)_R$, is tightly related to parity.
The eight components of the two Higgs multiplets are
in one-to-one correspondence with the eight scalar and pseudoscalar
bilinear quark operators that can be built with one generation of quarks.

A second work \cite{Machet2} will be devoted  to the mass spectrum of fermions,
gauge, Higgs and (pseudo)Goldstone bosons, to their couplings,
 and to the peculiar properties of the second Higgs boson $s$.
The case of $N$ generations will also be evoked; 
there exist in this case $2N^2$ Higgs multiplets isomorphic to the one of the GSW
model, which should {\em a priori} all be incorporated, like we did above
for $H$ and $K$; fermion mixing can then be taken care of
in a natural way in the embedding of the weak $SU(2)_L$ group
into the chiral group $U(2N)_L \times U(2N)_R$ \cite{Machet}.

\bigskip
{\em \underline{Acknowledgments:} it is a pleasure to thank
 O.~Babelon, M.~Capdequi-Peyran\`ere, S.~Davidson, M.~Knecht, J.~Lavalle,
 G.~Moultaka and M.I.~Vysotsky for conversations and advice.
They also greatly helped me to write
the present version of this work and of the one(s) to come.}


\begin{em}

\end{em}


\vskip .5cm

\appendix{\Large\bf Appendix}

\section{Expression of $\boldsymbol{SU(2)_L}$ and $\boldsymbol{SU(2)_R}$ generators
in the basis of the four components of $\boldsymbol K$ or
$\boldsymbol H$}
\label{section:4basis}

They are $4\times 4$ matrices which act on 4-vectors with basis
{\small
\begin{equation*}
h^0=\left(\begin{array}{c} 1\cr 0\cr 0\cr 0\end{array}\right),\ 
h^3=\left(\begin{array}{c} 0\cr 1\cr 0\cr 0\end{array}\right),\ 
h^+ = \left(\begin{array}{c} 0\cr 0\cr 1\cr 0\cr\end{array}\right),\ 
h^- = \left(\begin{array}{c} 0\cr 0\cr 0\cr
1\cr\end{array}\right).
\label{eq:4basis}
\end{equation*}
}
$\ast$\ The three $SU(2)_L$ generators write
{\small
\begin{equation*}
T^3_L = \left(\begin{array}{rrrr}
0 & -\frac12 & 0 & 0 \cr
-\frac12 & 0 & 0 & 0 \cr
0 & 0 & -\frac12 & 0 \cr
0 & 0 & 0 & \frac12 \end{array}\right),\quad
T^+_L = \left(\begin{array}{rrrr}
0 & 0 & 0 & -1 \cr
0 & 0 & 0 & -1 \cr
-\frac12 & \frac12 & 0 & 0 \cr
0 & 0 & 0 & 0 \end{array}\right),\quad
T^-_L = \left(\begin{array}{rrrr}
0 & 0 & -1 & 0 \cr
0 & 0 & 1 & 0 \cr
0 & 0 & 0 & 0 \cr
-\frac12 & -\frac12 & 0 & 0 \end{array}\right);
\label{eq:t3l}
\end{equation*} 
}
$\ast$\ the three $SU(2)_R$ generators write
{\small
\begin{equation*}
T^3_R = \left(\begin{array}{rrrr}
0 & \frac12 & 0 & 0 \cr
\frac12 & 0 & 0 & 0 \cr
0 & 0 & -\frac12 & 0 \cr
0 & 0 & 0 & \frac12 \end{array}\right),\quad
T^+_R = \left(\begin{array}{rrrr}
0 & 0 & 0 & 1 \cr
0 & 0 & 0 & -1 \cr
\frac12 & \frac12 & 0 & 0 \cr
0 & 0 & 0 & 0 \end{array}\right),\quad
T^-_R = \left(\begin{array}{rrrr}
0 & 0 & 1 & 0 \cr
0 & 0 & 1 & 0 \cr
0 & 0 & 0 & 0 \cr
\frac12 & -\frac12 & 0 & 0 \end{array}\right);
\label{eq:t3r}
\end{equation*} 
}
$\ast$\ the three $SU(2)_V$ generators $\vec T= \vec T_L + \vec T_R$ are
accordingly
{\small
\begin{equation*}
T^3 = \left(\begin{array}{rrrr}
0 & 0 & 0 & 0 \cr
0 & 0 & 0 & 0 \cr
0 & 0 & -1 & 0 \cr
0 & 0 & 0 & 1 \end{array}\right)= Q,\quad
T^+ = \left(\begin{array}{rrrr}
0 & 0 & 0 & 0 \cr
0 & 0 & 0 & -2 \cr
0 & 1 & 0 & 0 \cr
0 & 0 & 0 & 0 \end{array}\right),\quad
T^- = \left(\begin{array}{rrrr}
0 & 0 & 0 & 0 \cr
0 & 0 & 2 & 0 \cr
0 & 0 & 0 & 0 \cr
0 & -1 & 0 & 0 \end{array}\right),
\label{eq:t3}
\end{equation*} 
}

\end{document}